\documentclass[pra,10pt,eqsecnum,twocolumn]{revtex4}

\begin{document}

\newcommand{\sch}{Schr\"odinger }
\newcommand{\schs}{Schr\"odinger's }
\newcommand{\nn}{\nonumber}
\newcommand{\nl}{\nn \\ &&}
\newcommand{\dg}{^\dagger}
\newcommand{\bra}[1]{\langle{#1}|}
\newcommand{\ket}[1]{|{#1}\rangle}
\newcommand{\Bra}[1]{\Big{\langle}{#1}\Big{|}}
\newcommand{\Ket}[1]{\Big{|}{#1}\Big{\rangle}}
\newcommand{\bl}{{\Bigl(}}
\newcommand{\br}{{\Bigr)}}
\newcommand{\erf}[1]{Eq.~(\ref{#1})}
\newcommand{\erfs}[2]{Eqs.~(\ref{#1}) and (\ref{#2})}
\newcommand{\erft}[2]{Eqs.~(\ref{#1}) -- (\ref{#2})}
\newcommand{\beq}{\begin{equation}} 
\newcommand{\eeq}{\end{equation}}

\title{Comment on ``Testing integrability with a single bit of quantum information''}

\date{\today}
\author{H. M. Wiseman} 
\affiliation{Centre for Quantum Computer Technology, Centre for Quantum Dynamics, School of Science, Griffith University, Brisbane 4111 Australia.
email: h.wiseman@griffith.edu.au}

\begin{abstract}
In quant-ph/0303042, Poulin, Laflamme, Milburn and Paz consider the problem of 
distinguishing quantum chaos from quantum integrability for dynamics in an $N$-dimensional Hilbert space. They claim that this can be done by  
 deterministic quantum computing with a single bit  using $O(\sqrt{N})$ physical resources, compared to $O(N)$ physical resources classically. I 
 point out what seems to be a fatal flaw with their proposal.
 \end{abstract}

\maketitle
Ref.~\cite{Pou03} is concerned with the problem of determining whether the dynamics of a quantum system in an $N$-dimensional Hilbert space exhibit quantum chaos or quantum integrability. That is, whether the classical limit of the system is chaotic or integrable. 
This is a question of the properties of the distribution of the eigenvalue-spacing  of the $N\times N$ Hamiltonian matrix $H$. They claim that this problem be solved using 
 deterministic quantum computing with a single pseudo-pure qubit (DCQ1) 
 \cite{KniLaf98} using $O(\sqrt{N})$ physical resources. For example, repeating the experiment $O(\sqrt{N})$ times, or using $O(\sqrt{N})$ computers in parallel, as in NMR quantum computing (QC) \cite{Cor00}. This represents a quadratic speed up over a classical computer, which requires $O(N)$ resources to find the $N$ eigenvalues of $H$.
 
The crucial point of their proposal is at the end of their Sec.~III. 
For the quantum algorithm to work, it is necessary to distinguish 
a signal difference of order $1/\sqrt{N}$ for a single computer.
At this point they simply say ``... which can be achieved using $O(\sqrt{N})$ physical
 resources." Unfortunately, this seems not to be  the case.

In a single qubit, if the signal (the mean value of $\sigma_z$, say) is small (of order $1/\sqrt{N}$ here), then the noise in a measurement of $\sigma_z$ (``projectiion noise'') is of order unity. That is, the results $+1$ and $-1$ occur with almost equal probability so the variance in $\sigma_z$ is almost 1. Now the {\em fundamental}  signal to noise ratio (SNR) for an ensemble of such computers is proportional to 
$\sqrt{M}$, where $M$ is the size of the ensemble. This is true whether $M$ is the number of computers running in parallel, or the number of  repetitions for a single computer. This is an elementary  result in statistics: the signal rises as $M$ but the noise rises as $\sqrt{M}$.  
Now to see the signal requires that the SNR be of order unity. Therefore one requires $M$ of order $N$. That is, the resources scale the same as they do classically.
 
 For relatively small $M$, the practical SNR in NMR QC may scale as $M$. This is because the
 noise may be limited by technical noise rather than fundamental
 projection noise, so the SNR rises with the signal. But that makes no difference to my proof 
 because this argument is true only while technical noise is greater
 than projection noise. Thus the resources required can never be smaller than
 that required by projection noise, namely $O(N)$.

\end{document}